\documentclass[a4paper,12pt,abstract=true,enabledeprecatedfontcommands]{scrartcl}
\usepackage{authblk}

\usepackage{amsmath, amssymb}
\usepackage{tikz}
\usepackage[colorlinks=true,allcolors=blue]{hyperref}

\begin{document}
\titlehead{\hfill KUNS-2721,  UCRHEP-T590}

\title{\Large Syndetic Extension of Baryon and Lepton Numbers: \\
Proton Decay and Long-Lived Dark Matter}

\author[1]{Ernest~Ma}
\affil[1]{Physics and Astronomy Department, 
 University of California, Riverside, California 92521, USA} 

\author[2]{Koji~Tsumura\thanks{Email: \texttt{ko2@gauge.scphys.kyoto-u.ac.jp}}}
\affil[2]{Department of Physics, Kyoto University, Kyoto 606-8502, Japan}

\date{\normalsize May 10, 2018}

\maketitle

\begin{abstract}
The well-known baryon and lepton numbers of the standard model of quarks and 
leptons are extended to include new fermions and bosons in a simple structure 
with several essential features.  The usual heavy right-handed neutrino 
singlets (for neutrino mass and leptogenesis) are related to the axion 
which solves the strong CP problem.  At the same time, baryon number is 
broken softly, allowing the proton to decay.  Associated with this breaking, 
a long-lived dark-matter candidate (called the pseudo-sakharon) emerges.  
This new insight connects proton decay to a new component of dark matter.
\end{abstract}

\section{Introduction}

The standard model (SM) of quarks and leptons is known to have the built-in 
global $U(1)$ symmetries of baryon number $B$ and lepton number $L$.  If 
new particles are added, their $B$ and $L$ assignments may be chosen 
judiciously~\cite{m17} to address a number of outstanding theoretical 
issues.  In the following, it will be shown how a simple extension of the 
SM, which connects~\cite{s87,m01} the seesaw neutrino mass with the axion 
decay constant, may also allow a new understanding of the longevity~\cite{m13} 
of weak-scale dark matter (DM), i.e. that it is related to proton decay.

If the SM  is to be extended, one may want to consider the fundamental 
issues of (I) nonzero neutrino mass, (II) DM, and (III) strong 
$CP$ nonconservation in quantum chromodynamics (QCD). 
A simple connection was proposed thirty years ago~\cite{s87}, where  
the neutrino mass seesaw anchor scale is identified with the vacuum 
expectation value of a singlet scalar field which couples anomalously 
to new very heavy quark singlet fields $(\Psi_{L,R})$ as well as the three 
very heavy right-handed neutrinos $(N_R)$.  Together with the well-known 
mechanism of 
leptogenesis~\cite{fy86}, this also explains the baryon asymmetry of 
the Universe.  The DM of this model is the invisible axion 
which is yet to be discovered.  However, it is not guaranteed that 
the axion accounts for all of DM.  In fact, the anomalous 
Peccei-Quinn symmetry~\cite{pq77} which yields the axion~\cite{we78,wi78} 
has in general a residual discrete $Z_2$ symmetry~\cite{dmt14} which 
may be relevant for weak-scale DM.  In that case, the strong 
$CP$ problem may well be solved by the axion, but the latter may only be 
a small component of DM, whereas the bulk comes from  
a weak-scale DM particle, odd 
under this $Z_2$.   The axion-neutrino connection implies the basic 
assumption~\cite{mrz17,mot17} $U(1)_{PQ} = U(1)_L$, and that $\Psi$ transforms 
under both $B$ and $L$.

Another theoretical issue is whether or not DM is truly stable, 
in which case it should be protected by a symmetry, or just a very 
long-lived particle such as the invisible axion.  It is now known 
that such a DM particle must have a lifetime orders of 
magnitude longer than the age of the Universe, to avoid disrupting~\cite{sw17} 
the cosmic microwave background 
and other astrophysical 
observations~\cite{mpq16}.  The only possible exception is for the DM 
to decay dominantly to neutrinos, which was implemented in a recent 
model~\cite{mpsz15} where lepton number becomes a discrete $Z_3$ symmetry.

If DM is not absolutely stable, then its lifetime must be very 
long.  This may be due to the possible unification of matter and DM 
at a very high scale~\cite{m13} or that it is somehow related to 
a known lifetime which is very long.  This brings to mind proton decay 
and in this paper it will be shown how the two may be related.

This paper is organized as follows. 
In Section 2, the particle content and the relevant interactions are introduced, 
which define the unique lepton and baryon numbers of the new particles.  
In Section 3, the longevity of the DM is linked to the proton decay. 
A scenario for thermal freeze out of the DM is also discussed. 
Summary of our new proposal is given in Section 4. 

\section{The Model}

The axion-neutrino connection is established using a very heavy colored electroweak singlet quark 
$\Psi$ and three very heavy right-handed singlet neutrinos, as shown in Table~1.
\begin{table}[tbh]
\centering
\begin{tabular}{|c|c|c|c|}
\hline 
Particle & $SU(3)_C \times SU(2)_L \times U(1)_Y$ & $B$ & $L$ \\ 
\hline \hline
$q_{L}^{} = (V_{\text{CKM}}^{\dag}u,d)_L^{}$ & $(3, 2, 1/6)$ & $1/3$ & $0$ \\
$u_R^{}$ & $(3, 1, 2/3)$ & $1/3$ & $0$ \\ 
$d_R^{}$ & $(3, 1, -1/3)$ & $1/3$ & $0$ \\ 
\hline
$\Psi_L$ & $(3, 1, -1/3)$ & $-2/3$ & $0$ \\ 
$\Psi_R$ & $(3, 1, -1/3)$ & $-2/3$ & $-1$ \\ 
\hline
$\ell_{L}^{} = (\nu,e)_L^{}$ & $(1, 2, -1/2)$ & $0$ & $1$ \\ 
$e_R^{}$ & $(1, 1, -1)$ & $0$ & $1$ \\
$N_R$ & $(1, 1, 0)$ & $0$ & $1$ \\ 
\hline
\hline
$\Phi=(\phi^+,\phi^0)$ & $(1, 2, 1/2)$ & $0$ & $0$ \\ 
\hline
$S_1$ & $(1, 1, 0)$ & $0$ & $-1$ \\ 
$S_2$ & $(1, 1, 0)$ & $0$ & $-2$ \\ 
$\zeta$ & $(3, 1, -1/3)$ & $-2/3$ & 0 \\
$\sigma$ & $(1, 1, 0)$ & $1$ & $0$ \\
\hline
\end{tabular}
\caption{Particle content of model with axion and pseudo-sakharon.}
\end{table}
Whereas $N_R$ has $L=1$ as usual, $\Psi$ is assumed to have $B=-2/3$ with 
$\Psi_{L,R}$ having $L=0,-1$.  Now the scalar singlets $S_1, S_2$ have 
$L=-1,-2$.  Hence the terms
\begin{align}
-{\mathcal L}_{S}^{}
&=
+ y_{\Psi}^{}\, S_1^{\star} \overline{\Psi_L} \Psi_R 
+ \frac12 y_N^{i}\, S_2 \overline{N_{iR}^{c}} N_{iR} 
+\kappa \, S_2^\star (S_1)^2 
+ \text{H.c.} \label{Eq:LS}
\end{align}
are allowed.  The $y_\Psi^{}$ term means that $U(1)_L = U(1)_{PQ}$ and its 
spontaneous breaking through the vacuum expectation value 
$\langle S_1 \rangle \neq 0$ generates the mass of the heavy quark, $M_\Psi$, 
as well as the QCD axion because $\Psi$ is a colored 
fermion.  The resulting axion particle is thus of the KSVZ 
type~\cite{k79,svz80}, 
and the domain wall number is $1$, so it is cosmologically safe~\cite{s82}. 
The $y_N^{}$ term means that $\langle S_2 \rangle$ generates 
the mass of right-handed neutirnos, $M_N$,  
but the would-be singlet majoron~\cite{cmp81,gr81} is now related to 
the axion~\cite{s87} through the $\kappa$ term. 
This simple idea says that both the neutrino seesaw 
anchor scale and the axion decay constant have a common origin, so 
the existence of one is tied to that of the other.  

The key of the present new model is the addition of $\zeta$ and $\sigma$.  
With only $\zeta$, the allowed terms are
\begin{align}
-{\mathcal L}_{\zeta}
&=
+ y^{ij}_L\, \zeta^{\star}\, \overline{q_{iL}^{}}\,i\tau^{2} q_{jL}^{c}
+ y^{ij}_R\, \zeta\, \overline{d_{iR}^{c}} u_{jR}^{} 
+ y^{i}_{\zeta}\, \zeta^\star\, \overline{{N}_{iR}^{c}} \Psi_R 
+ \text{H.c.}  
\end{align}
These terms justify the assignment that $\Psi$ has $B=-2/3$, and the model at 
this point conserves $B$.  In previous work~\cite{m08,m09}, 
it was shown how $B$ may be broken to $(-1)^{3B}$.  Here with the addition 
of $\sigma$, so that the term
\begin{align}
-{\mathcal L}_{\sigma}
&=
+ y^i_\sigma\, \sigma^{\star}\, \overline{\Psi_{L}} d_{iR}^{} + \text{H.c.} 
\end{align}
is allowed, $B$ is broken by $\langle \sigma \rangle = v_{\sigma} \neq 0$.\footnote{  
The idea that a scalar singlet carrying baryon number 
may have a vacuum expectation value was first proposed~\cite{m88} many years 
ago in the context of superstring-inspired $E_6$ models.}

With the spontaneous breaking of $B$ by $\langle \sigma \rangle$, a massless 
Nambu-Goldstone (NG) boson will appear.  It may be called the `sakharon'~\cite{m17} after 
Andrei Sakharov~\cite{s67}.  Such a massless particle coupled to baryon 
number would be highly constrained experimentally.  In this proposal, the 
soft term
\begin{align}
V_\text{soft} = -\frac{\mu^2}2 \sigma^2 + \rm{H.c.}
\end{align}
is added, which violates $B$ by two units, resulting in a massive pseudo-sakharon instead. 

For definiteness of the model parameters, we introduce the neutrino Yukawa couplings, 
which realize the conventional seesaw mechanism, i.e. 
\begin{align}
{\mathcal L}_{D}^{}
&= - y_{D}^{\ell j} \overline{L}_{\ell} \widetilde{\Phi} N_{j R} +\text{H.c.} 
\end{align}

For later convenience, we also define the mixing between the down-type quarks and the heavy quark $\Psi$ 
through the mass matrix linking them, i.e.
\begin{align}
\begin{pmatrix} \overline{d_{i L}^{}} & \overline{\Psi_{L}} \end{pmatrix}
\begin{pmatrix} m_{d}^{i} & 0 \\ y_{\sigma}^{i}v_{\sigma}^{} & M_{\Psi} \end{pmatrix}
\begin{pmatrix} d_{i R}^{} \\ \Psi_{R} \end{pmatrix}
\to 
\begin{pmatrix} \overline{d_{i L}^{}} & \overline{\Psi_{L}} \end{pmatrix} U_{L}^{\dag}
\begin{pmatrix} \widehat{m_{d}}^{i} & 0 \\ 0 & \widehat{M_{\Psi}} \end{pmatrix}
U_{R} \begin{pmatrix} d_{i R}^{} \\ \Psi_{R} \end{pmatrix}, 
\end{align}
where
\begin{align}
U_{X} (X=L,R)=
\begin{pmatrix} \cos\theta_{X}^{} & \sin\theta_{X}^{} \\ -\sin\theta_{X}^{}& \cos\theta_{X}^{} \end{pmatrix}. 
\end{align}
We then rename the $d$ and $\Psi$ fields as the ones in the basis of their mass eigenstates. 
The resulting right-handed mixing is approximately given by $\theta_{R}^{} \approx y_{\sigma}v_{\sigma}/M_{\Psi}^{}$, while 
the left-handed mixing is further suppressed, i.e. $\theta_{L}^{} \approx m_{d}y_{\sigma}v_{\sigma}/M_{\Psi}^{2}$. 
\\


The scalar potential consisting of $\Phi$, $S_{1}$, $S_{2}$, and $\sigma$ 
is simply given by
\begin{align}
V 
&= 
-\mu_\Phi^2 \Phi^\dagger \Phi 
-\mu_1^2 |S_1|^{2} -\mu_2^2 |S_2|^{2} 
-\mu_\sigma^2 |\sigma|^2 
- \frac{\mu_{}^2}2 (\sigma^2 + \text{H.c.}) 
-\kappa \big( S_2^\star (S_1)^2 + \text{H.c.} \big) \nonumber \\ 
& \qquad 
+ \frac{\lambda_{\Phi}^{}}2 (\Phi^\dagger \Phi)^2 
+ \frac{\lambda_1^{}}2 |S_1|^4 + \frac{\lambda_2^{}}2 |S_2|^{4} 
+ \frac{\lambda_\sigma^{}}2 |\sigma|^{4} 
+\lambda_{12} |S_1|^{2}|S_2|^{2}
\nonumber \\ 
& \qquad 
+ ( \lambda_{1\Phi}^{} |S_1|^{2} + \lambda_{2\Phi}^{} |S_2|^{2}
+ \lambda_{\Phi\sigma}^{} |\sigma|^{2}) (\Phi^\dagger \Phi)
+ ( \lambda_{1\sigma} |S_1|^{2} + \lambda_{2\sigma} |S_2|^{2} ) |\sigma|^{2}.   
\end{align}
In addition to being invariant under the SM gauge symmetry, it is also 
invariant under $U(1)_L$ and $U(1)_B$ except for the soft $\mu^2$ term 
which breaks $U(1)_B$ to the baryon triality, $(-1)^{3B}$.
No other soft breaking term such as $\sigma, \sigma^{3}, (\Phi^{\dag}\Phi)\sigma,$ etc. is introduced 
since these terms disturb the observed baryon triality relation. 
Let 
$\langle \phi^0 \rangle = v = 174\,\text{GeV}$ which breaks $SU(2)_L \times U(1)_Y$ to 
$U(1)_Q$, $\langle S_1 \rangle = f_1^{}$ and $\langle S_2 \rangle = f_2^{}$ 
which break $U(1)_L$, and  $\langle \sigma \rangle = v_\sigma$ which breaks 
$U(1)_B$, then the minimum of $V$ is determined by
\begin{align}
&-\mu_\Phi^2 +\lambda_{\Phi}^{} v_{}^{2}
+\lambda_{1\Phi}^{} f_{1}^{2} + \lambda_{2\Phi}^{} f_{2}^{2} +\lambda_{\Phi\sigma}^{} v_{\sigma}^{2} 
=0, \\
&-\mu_1^2 - 2\, \kappa\, f_{2}^{} + \lambda_1 f_{1}^{2}
+\lambda_{12} f_{2}^{2} +\lambda_{1\Phi}^{} v_{}^{2} +\lambda_{1\sigma} v_{\sigma}^{2} 
=0, \\
&-\mu_2^2 -\kappa\, \big( f_{1}^2/f_{2}^{} \big) +\lambda_2 f_{2}^{2} 
+\lambda_{12} f_{1}^{2} +\lambda_{2\Phi}^{} v_{}^{2} +\lambda_{2\sigma} v_{\sigma}^{2}
=0, \\
&-\mu_\sigma^2  -\mu^2 +\lambda_\sigma v_{\sigma}^{2} 
+\lambda_{\Phi\sigma}^{} v_{}^{2} +\lambda_{1\sigma} f_{1}^{2} +\lambda_{2\sigma} f_{2}^{2}
=0. 
\end{align}
Let
\begin{align}
\Phi &= \begin{pmatrix} i\, \eta^{+} \\ v_{} +\frac{\rho_{}^{}+i\, \eta_{}^{}}{\sqrt2} \end{pmatrix},~ 
S_1 = f_1 + \frac{\rho_1^{} + i\, \eta_1^{}}{\sqrt2},~ 
S_2 = f_2 + \frac{\rho_2^{} + i\, \eta_2^{}}{\sqrt2},~ 
\sigma = v_{\sigma} +\frac{\rho_\sigma^{} + i\, \eta_\sigma^{}}{\sqrt2}. 
\end{align}
then the $2 \times 2$ mass-squared matrix spanning $(\eta_1,\eta_2)$ is 
given by
\begin{align}
{\mathcal M}_{\eta}^{2}
&= \kappa \begin{pmatrix} 4 f_{2}^{} & - 2 f_{1}^{} \\ -2 f_{1}^{} & f_{1}^2/f_{2}^{} \end{pmatrix}.   
\end{align}
The massless NG boson mode corresponds to the spontaneous breaking 
of $U(1)_L$, i.e. the majoron $J$. 
We assume $f_{1, 2}^{} \gg v, v_\sigma$, hence $f = \sqrt{f_1^2 + 4f_2^2}$ is the axion 
decay constant $F_A$ and must be large~\cite{rs88}: $F_A > 4 \times 10^8$ 
GeV, and $J$ is also the QCD axion.  Also from Eq.\eqref{Eq:LS}, $f_2^{}$ 
determines the neutrino seesaw anchor scale.  The decays of the lightest 
$N$ generate a lepton asymmetry which gets converted by sphalerons to 
the present baryon asymmetry of the Universe.
The state $K$ orthogonal to $J$ is assumed to be superheavy with 
$M_{}^2 = \kappa(f_1^2 + 4f_2^2)/f_2$:
\begin{align}
\begin{pmatrix} J \\ K \end{pmatrix}
=
\begin{pmatrix} f_{1}^{}/f & 2f_{2}^{}/f \\ -2f_{2}^{}/f & f_{1}^{}/f \end{pmatrix}
\begin{pmatrix} \eta_{1}^{} \\ \eta_{2}^{} \end{pmatrix}. 
\end{align}
%
A pseudo-NG boson associated with the baryon number, 
the pseudo-sakharon $S \equiv \eta_{\sigma}^{}$, is also generated with mass given by 
\begin{align}
m_{S}^{2} 
&= 
2\mu^{2}_{}. 
\end{align}
Note that $\mu_{}^{2}$ is the soft breaking term of the baryon number conservation. 
The pseudo-sakharon $S$ is the (long-lived) DM candidate in this model. 
The mass matrices for real components of the fields spanning $(\rho_{1}^{}, \rho_{2}^{})$ 
and $(\rho, \rho_{\sigma}^{})$  are 
\begin{align}
{\mathcal M}_{\rho}^{2}
&= \begin{pmatrix} 2 \lambda_{1}f_{1}^{2} & -2 \kappa f_{1}^{} \\ 
-2 \kappa f_{1}^{} &  2 \lambda_{2}f_{2}^{2} + \kappa \big(f_{1}^{2}/f_{2}^{}\big) \end{pmatrix},  \quad
{\mathcal M}_{}^{2}
= \begin{pmatrix} 2 \lambda_{\Phi}^{} v_{}^{2} & 2 \lambda_{\Phi\sigma}^{} v\, v_{\sigma} \\ 
2 \lambda_{\Phi\sigma}^{} v\, v_{\sigma} &  2 \lambda_{\sigma} v_{\sigma}^{2} \end{pmatrix}. 
\end{align}
The mass eigenstates are defined as 
\begin{align}
\begin{pmatrix} h_{125} \\ h_{\sigma} \end{pmatrix}
=
\begin{pmatrix} \cos\theta_{h}^{} & \sin\theta_{h}^{} \\ -\sin\theta_{h}^{}& \cos\theta_{h}^{} \end{pmatrix}
\begin{pmatrix} \rho_{}^{} \\ \rho_{\sigma}^{} \end{pmatrix}. 
\end{align}
where $\tan2\theta_{h}^{} 
=2\lambda_{\Phi\sigma}^{}vv_{\sigma}/(\lambda_{\Phi}^{}v^{2}-\lambda_{\sigma}^{}v_{\sigma}^{2})$. 
Hereafter, we assume $\lambda_{\Phi\sigma} \ll 1$ in order to avoid the stringent constraints from 
the Higgs invisible decay; 
\begin{align}
\Gamma(h_{125}\to SS)
&\approx \frac1{2!} \frac{(\sqrt2 \lambda_{\Phi\sigma}^{}v)^{2}}{16\pi m_{h}^{}} 
\sqrt{1-\frac{4m_{S}^{2}}{m_{h}^{2}}},  
\end{align}
and the DM direct detection search; 
\begin{align}
\sigma_{DD}^{}
&= \lambda_{\Phi\sigma}^{2} \frac{f_{N}^{2}}{16\pi} \frac{\overline{m}^{2}m_{N}^{2}}{m_{S}^{2}m_{h}^{4}}, 
\end{align}
where $\overline{m} = m_{S}^{}m_{N}^{}/(m_{S}^{}+m_{N}^{})$ and $f_{N}^{}=0.308 \pm 0.018$. 
Thus, $\theta_{h}^{} \approx 2\lambda_{\Phi\sigma}^{}vv_{\sigma}/(m_{h}^{2}-m_{\sigma}^{2})$, where 
the masses of these scalar bosons are approximately given by 
$m_{h_{125}}^{2} \equiv m_{h}^{2} \approx 2\lambda_{\Phi}^{}v^{2}$ 
and $m_{h_{\sigma}}^{2} \equiv m_{\sigma}^{2} \approx 2\lambda_{\sigma}v_{\sigma}^{2}$. 
%

\section{Long-Lived Dark Matter and Proton Decay}

The pseudo-sakharon $S$ can decay to $d\overline{d}$ only through the SM-heavy quark mixing 
$\theta_{L}^{} \approx m_{d}^{i}\, y_{\sigma}^{i} v_{\sigma}^{}/M_{\Psi}^{2}$ with rate given by 
\begin{align}
\Gamma(S \to d\overline{d}) 
\approx \frac{m_{S}^{}}{16\pi} (y_{\sigma}^{d})^{4} \Big( \frac{m_{d}^{} v_{\sigma}^{}}{M_{\Psi}^{2}} \Big)^{2} 
\sqrt{1-\frac{4m_{d}^{2}}{m_{S}^{2}}}. 
\end{align}
Thus, the longevity of the DM can always be maintained by choosing small $y_{\sigma}^{}$ as  in 
\begin{align}
\tau_{S}^{}
&\simeq 10^{27} \text{sec} 
\times 
\Big( \frac{M_{\Psi}^{}}{10^{10}\, \text{GeV}} \frac{6.5\times 10^{-3}}{y_{\sigma}^{d}} \Big)^{4}
\Big( \frac{4.7\, \text{MeV}}{m_{d}^{}} \frac{20\, \text{GeV}}{v_{\sigma}^{}} \Big)^{2}
\Big(\frac{20\, \text{GeV}}{m_{S}^{}}\Big). 
\label{Eq:Relic}
\end{align}
This is easily set to be greater than $10^{27}$ seconds to avoid all possible 
cosmological constraints on $S$ as a DM candidate. 
Similarly, $y^s_\sigma$ and $y^b_\sigma$ must be suppressed by the additional factor 
of $\sqrt{m_d/m_s}$ and $\sqrt{m_d/m_b}$ respectively.

For $S$ to be DM, the coupling $\lambda_{\Phi\sigma}$ must be small 
to satisfy the Higgs invisible decay~\cite{atlas:hinv,cms:hinv} 
 and the direct-search constraints~\cite{xenon17}. 
This means that the annihilation cross section of $S$ through 
the SM Higgs boson to SM particles is much smaller than the canonical value 
of $\sigma v_{\text{rel}} \approx 1$~pb to have the correct relic abundance.  
However, if $m_{S}^{} > m_{\sigma}^{}$,  
the relic abundance is fixed by the $SS \to h_{\sigma}^{}h_{\sigma}^{}$ process 
instead, at temperatures below $2m_{S}^{}$ but above $2m_{\sigma}^{}$. 
The annihilation cross section of $SS \to h_{\sigma}^{}h_{\sigma}^{}$ times its relative velocity is given by 
\begin{align}
\sigma v_{\rm{rel}} 
\approx \frac1{2!} \frac{\lambda_{\sigma}^{2}}{32\pi m_{S}^{2} }
\Big( 1 -\frac{3\,\xi}{4-\xi} +\frac{\xi}{2-\xi} \Big)^{2}\sqrt{1-\xi}, 
\end{align}
where $\xi=m_{\sigma}^{2}/m_{S}^{2}$. 
Note that the cross section is maximized for $\xi=0$, and vanishes for $\xi=1$. 
The pseudo-sakharon remains in thermal equilibrium with the SM particles through its scattering 
with $h_{\sigma}$ which in turn interacts and mixes with the SM Higgs and has effective Yukawa 
couplings to the SM fermions.  
As the temperature of the Universe drops below $2m_{S}^{}$, $S$ freezes out because 
it is effectively stable due to its very long lifetime.  
As for $h_{\sigma}$, it decays quickly away 
so that only SM particles remain in thermal equilibrium, and the usual 
big bang nucleosynthesis is not disturbed.    Note the crucial 
built-in condition that $S$ does not mix with the SM Higgs because 
of automatic $CP$ invariance in the scalar sector.

\begin{figure}[tb]
\centering 
\includegraphics[scale=0.6]{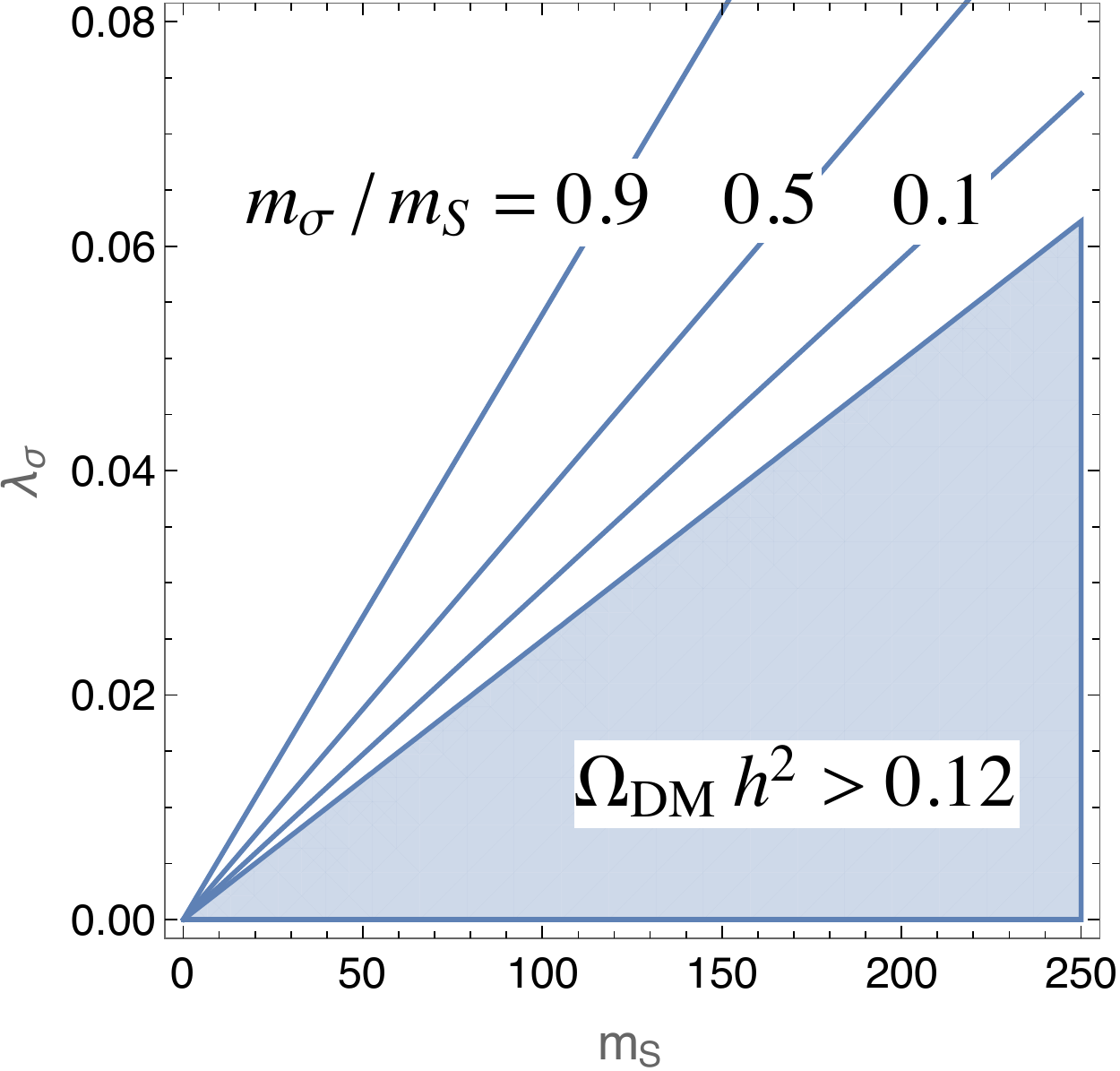}
\caption{Relic abundance constraints on $m_S$ and $\lambda_\sigma$ for different 
$m_\sigma/m_S$ values.}
\label{Fig:Relic}
\end{figure}

In Fig.\ref{Fig:Relic}, the constraint from obtaining 100\% of the DM relic 
abundance is given in the  
$(m_{S}^{}, \lambda_{\sigma}^{})$ plane for different values of $m_\sigma/m_S$. 
Within the parameter space of $m_S^{}$, $\lambda_\sigma$ and  $m_{\sigma}$ 
under the condition $m_{\sigma}^{2}\approx 2\lambda_{\sigma}v_{\sigma}^{2}$,  
there are clearly other allowed values.  
%
\\


\begin{figure}[tb]
\centering 
\includegraphics[scale=0.9]{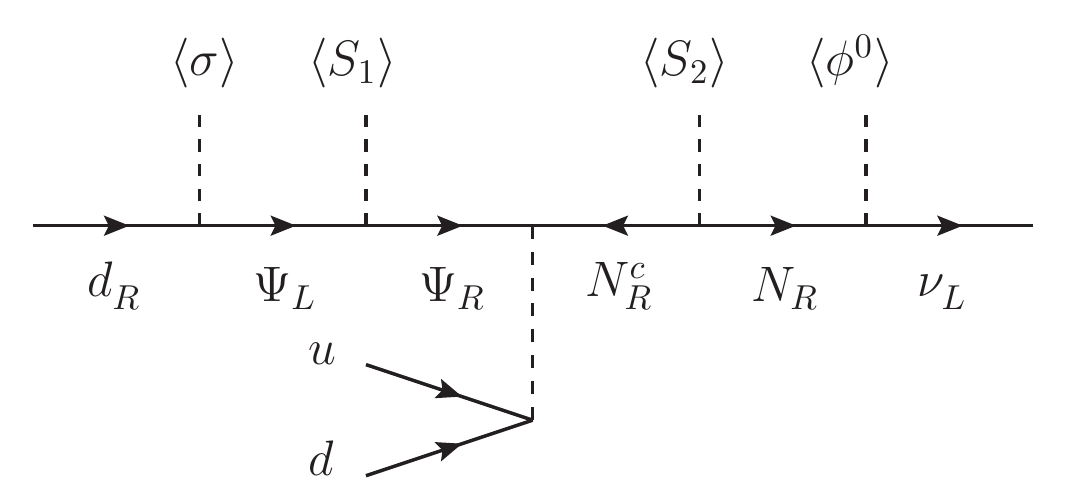}
\caption{Proton decay induced by $\Psi$, $N_{R}$, and $\zeta$.}
\label{Fig:PD}
\end{figure}

At energy scales below the new heavy particle masses $(M_{\zeta}, M_{\Psi}, M_{N})$, 
the following higher dimensional operators are generated;
\begin{align}
{\mathcal L}_{\text{dim} 6}^\text{eff}
&=
\frac{2}{M_{\zeta}^{2}} \Big| + y^{ij}_L\, \overline{q_{iL}^{}}\,i\tau^{2} q_{Lj}^{c}
+ y^{ij\star}_R\, \overline{u_{jR}^{}} d_{iR}^{c} \Big|^{2}, \\
{\mathcal L}_{\text{dim} 7}^\text{eff}
&=
 \frac{1}{M_{\zeta}^{2}} \frac{y_{\zeta}^{i} (y_{\sigma}^{i} v_{\sigma}^{}) y_{D}^{\ell i}}{M_{N}^{i} M_{\Psi}} 
\overline{\ell_{L}^{}} \widetilde{\Phi} d_{iR}^{}
\Big( + y^{ij \star}_L\, \overline{q_{iL}^{c}}\,i\tau^{2} q_{Lj}^{} + y^{ij}_R\, \overline{u_{jR}^{c}} d_{iR}^{} \Big) 
+ \text{H.c.} \\
{\mathcal L}_{\text{dim} 8}^\text{eff}
&=
\frac{2}{M_{\zeta}^{2}} \Big| \frac{y_{\zeta}^{i} (y_{\sigma}^{i} v_{\sigma}^{}) y_{D}^{\ell i}}{M_{N}^{i} M_{\Psi}} 
\overline{\ell_{L}^{}} \widetilde{\Phi} d_{iR}^{} \Big|^{2}. 
\end{align}

The dimension-six four quark operators are constrained by the LHC data. 
Parametrizing the coefficient of the dimension-six operators by $(2\pi)/\Lambda^{2}$, 
the lower bound on this contact interactions is about $12.8\,(17.5)\,$TeV depending on 
the sign of the operators\cite{Sirunyan:2018wcm}. 
This bound is easily evaded by choosing heavier diquark with smaller Yukawa couplings. 
Although the dimension-six operators are the new source of the quark contact interactions 
at the tree level, there are no tree-level contributions to meson mixing. 

Although both baryon number and lepton numbers are broken,  
the dimension-seven operators induce the $B+L$ conserving proton decay in Fig.~\ref{Fig:PD} 
as in some previous proposals~\cite{v95,bm12,gms16}. 
The dominant decay is $p \to \pi^+ \nu$ and not the usual $p \to \pi^0 e^+$, 
where the latter conserves $B-L$. 
Since the Higgs field is replaced by its VEV at low energy, the effective operators 
relevant for the proton decay are given by 
\begin{align}
{\mathcal L}_{p\to \pi^{+}\nu}
&=
C_{L} {\mathcal O}_{L} + C_{R} {\mathcal O}_{R} + \text{H.c.} 
\end{align}
where
\begin{align}
{\mathcal O}_{L} 
&= (\overline{q_{iL}^{c}}\,i\tau^{2} q_{Lj}^{})(\overline{\nu_{L}}d_{R}), 
&&{\mathcal O}_{R}
= (\overline{u_{jR}^{c}} d_{iR}^{})(\overline{\nu_{L}}d_{R}), \\
C_{L}
&= - \frac{1}{M_{\zeta}^{2}} \frac{y_{\zeta}^{i} (y_{\sigma}^{i} v_{\sigma}^{}) (y_{D}^{\ell i}v)}{M_{N}^{i} M_{\Psi}} 
y_{L}^{ij\star}, 
&& C_{R}
= - \frac{1}{M_{\zeta}^{2}} \frac{y_{\zeta}^{i} (y_{\sigma}^{i} v_{\sigma}^{}) (y_{D}^{\ell i}v)}{M_{N}^{i} M_{\Psi}} 
y_{R}^{ij}.  
\end{align}
The proton decay rate is calculated from this effective Lagrangian as 
\begin{align}
\Gamma(p\to\pi^{+}\nu)
= \frac{m_{p}}{32\pi} \Big( 1-\frac{m_{\pi}^{2}}{m_{p}^{2}} \Big)^{2} 
\Big| C_{L} \langle \pi^{+}|(ud)_{L}^{}d_{R}^{}|p\rangle 
+ 2C_{R} \langle \pi^{+}|(ud)_{R}^{}d_{R}^{}|p\rangle \Big|^{2}. 
\end{align}
Thanks to the parity symmetry in QCD, 
$\langle \pi^{+}|(ud)_{\Gamma}^{}d_{R}^{}|p\rangle=\langle \pi^{+}|(ud)_{\Gamma}^{}d_{L}^{}|p\rangle (\simeq 0.18)$, 
which are given in Ref.\cite{Aoki:2017puj}. 
Then the proton lifetime is evaluated as 
\begin{align}
\tau_{p\to \pi^{+}\nu} 
&\simeq 8 \times 10^{32}\, \text{yr} \times 
\Big( \frac{M_{\zeta}}{3\, \text{TeV}} \Big)^{4}
\Big( \frac{M_{\Psi}}{10^{10} \text{GeV}} \frac{20\text{GeV}}{v_{\sigma}} \Big)^{2}
\Big( \frac{M_{N}}{10^{10} \text{GeV}} 
\frac{10^{-10} \text{GeV}}{m_{\nu}} \Big) \nonumber \\
& \quad \times 
\Big( \frac{0.1}{y_{\zeta}} \,
\frac{6.5\times10^{-3}}{y_{\sigma}} \Big)^{2}
\Big( \frac{y_{L}^{}}{10^{-2}} \frac{\langle \pi^{+}|(ud)_{L}^{}d_{R}^{}|p\rangle}{0.18}
+2 \frac{y_{R}^{}}{10^{-2}} \frac{\langle \pi^{+}|(ud)_{R}^{}d_{R}^{}|p\rangle}{0.18}\Big)^{-2}. 
\end{align}
The current lower limit on this mode would be the same as that of $B-L$ conserving decay mode, 
i.e., $\tau_{p\to \pi^{+}\bar{\nu}} > 3.9 \times 10^{32}\, \text{yr}$\cite{Abe:2013lua}. 
The observable proton decay may be within reach in future experiments. 
Note that the longevity of the proton is now linked to the longevity of DM 
and also the smallness of the neutrino mass. 

\section{Summary}

We have constructed a model which connects the proton and DM longevity and 
the smallness of neutrino mass. 
In our new proposal, 
the lepton number symmetry ($L$) is identified as the PQ symmetry in the KSVZ model 
together with the conventional seesaw mechanism, where the pseudo-NG boson associated 
with the lepton number symmetry breaking behaves as the QCD axion. 
We have defined the lepton number of the heavy colored fermion $\Psi$ by introducing 
 two new scalar fields $S_{1}$ and $S_{2}$ whose lepton numbers are different. 
At the same time, the uniquely defined baryon number ($B$) is assigned to $\Psi$ through the diquark $\zeta$. 
We then introduced a new scalar $\sigma$ charged under $B$.  
The spontaneous breaking of $B$ by $\langle \sigma \rangle$ as well as an 
explicit soft violating term result in  
the pseudo-NG boson, dubbed the pseudo-sakharon, which is identified as the new long-lived DM, with $B$ broken to $(-1)^{3B}$. 
 Consequently, this model predicts the dominant $B+L$ conserving proton decay, i.e., $p\to \pi^{+}\nu$ and not the usual $p \to \pi^0 e^+$ which conserves 
$B-L$. 
 This new connection between the DM longevity and the proton longevity 
 opens up a new understanding of possible long-lived DM.

\section*{Acknowledgments}
This work was supported in part 
by the U.~S.~Department of Energy Grant No. DE-SC0008541 (EA), 
and by JSPS KAKENHI Grant Number 16K17697 (KT).  EM thanks Kyoto 
University for support during a recent visit.


\end{document}